\documentclass[usenatbib,twocolumns]{mnras}
\usepackage{graphicx}
\usepackage{amsmath}
\usepackage{textgreek}
\usepackage{amssymb}
\usepackage{bm}
\usepackage{color}
\usepackage{accents}
\usepackage{physics}
\usepackage{subfigure}
\usepackage{natbib}

\usepackage{yfonts}

\def\apj{ApJ}
\def\apjs{ApJS}
\def\aap{A \& Ap}

\def\araa{ARAA}

\def\mnras{MNRAS}
\def\apj{ApJ}
\def\apjl{ApJL}

\def\aap{A\&A}

\def\pre{Physical Review E}

\newcommand{\beq}{\begin{equation}}
\newcommand{\eeq}{\end{equation}}

\newcommand{\benum}{\begin{enumerate}}
\newcommand{\eenum}{\end{enumerate}}

\def\bar#1{\overline{#1}}
\newcommand{\del}{{\bm{\nabla}}}

\newcommand{\bmB}{{\bm{B}}}

\def\bmu{\bm{u}}

\newcommand{\emf}{{\mathcal{E}}}

\title[How to calculate turbulent transport coefficients]
{Calculating turbulent transport tensors by averaging single plume dynamics and application to  dynamos}

\author[Zhou \& Blackman]
{Hongzhe Zhou $^{1,2}$
\thanks{Email address for correspondence: hzhou21@ur.rochester.edu},
Eric G. Blackman $^{1,2}$
\thanks{Email address for correspondence: blackman@pas.rochester.edu}\\
$^1$ Department of Physics and Astronomy, University of Rochester, Rochester, NY, 14627, USA\\
$^2$ Laboratory for Laser Energetics,  University of Rochester, Rochester NY, 14623, USA\\
}

\begin{document}

\date{}
\pagerange{\pageref{firstpage}--\pageref{lastpage}} \pubyear{}
\maketitle
\label{firstpage}

\begin{abstract}
Transport coefficients in turbulence are comprised of correlation functions between turbulent fluctuations and efficient methods to calculate them are desirable. 
For example, in  mean field dynamo theories used to model the growth of large scale magnetic fields of  stars and galaxies,  the turbulent electromotive force is commonly approximated by a series of tensor products of turbulent transport coefficients with  successively higher order spatial derivatives of the mean magnetic field. One  ingredient of standard models  is the kinematic  coefficient of the zeroth order term, namely the averaged kinetic pseudotensor  $\bm\alpha$, that converts toroidal to poloidal fields.
Here we demonstrate  an efficient way to  calculate this quantity for rotating stratified turbulence, whereby the pre-averaged quantity is calculated for the motion of a single plume, and the average is then taken over an ensemble  of plumes of different orientations.
 We calculate the plume dynamics in the most convenient  frame, before transforming back to the lab frame and averaging. Our concise  configuration space calculation 
gives essentially identical results to   previous  lengthier  approaches. 
The present  application exemplifies what is a   broadly applicable method.
\end{abstract}

\begin{keywords}
dynamo - stars: magnetic fields - galaxies: magnetic fields - turbulence - methods: analytical - MHD
\end{keywords}

\section{Introduction}

Correlation functions of turbulent fluctuations in fluid dynamics and magnetohydrodynamics are needed to characterize mean transport coefficients of turbulent systems.
Examples include turbulent diffusion, Reynolds and Maxwell stresses, and kinetic, current, and cross helicities.
However, computing these quantities in anisotropic, stratified rotators of astrophysics can be cumbersome. 
Here we demonstrate a method which simplifies  this task, applying it to 
 the helical pseudotensor  of large scale dynamo heory as an example.

The context of mean field dynamo arises because magnetic fields are ubiquitous in  stars and galaxies whose ages are usually much larger than the magnetic diffusion time scales.  
Consequently, a reliable mechanism is necessary for the {\it in situ} generation and sustenance of magnetic fields.
In the framework of magnetohydrodynamics (MHD),  dynamo theory describes the conversion of flow kinetic energy into magnetic energy, and mean field dynamo theory specifically concerns  the magnetic fields which have a correlation scale larger than the turbulent scale.   In mean field theory, the induction equation is suitably averaged and the correlations between the fluctuating velocity and magnetic fields lead to an turbulent electromotive force (EMF)
\beq
\bm\emf=\bar{\bmu\times\bm b},
\label{eqn:emf}
\eeq
whose curl appears as a source term in the time evolution of the mean magnetic field.
In Eqn. (\ref{eqn:emf}), $\bmu$ and $\bm b$ are respectively the turbulent velocity and magnetic fields, i.e., $\bmu=\bm U_\text{tot}-\bar{\bm U}$ and $\bm b=\bm B_\text{tot}-\bar{\bm B}$, with $\bm U_\text{tot}$ being the total velocity field, $\bm B_\text{tot}$ the total magnetic field, and $\bar{\bm U}$ and $\bar{\bm B}$ the corresponding mean fields.
The overline indicates some appropriate mean which satisfies Reynolds rules;
hereafter we will drop the overlines for $\bar{\bm U}$ and $\bar{\bm B}$ for simplicity.

The turbulent EMF is commonly expanded in powers of the gradients of $\bmB$ whose spatial scale is assumed to be much larger than the scale of the turbulent fields, i.e.,
\beq
\emf_i=\alpha_{ij}B_j-\beta_{ijk}\partial_jB_k+\cdots,
\eeq
where $\bm\alpha$, $\bm\beta$, ..., are the turbulent transport coefficients.
determined by turbulent velocity and magnetic fields.
However, even without including the back-reaction from magnetic fields on the flow, finding robust analytical expressions for the coefficients is not  easy.
Common approaches  involve integrating the equations of motion of $\bmu$ and $\bm b$, combined with a closure.
For example, \cite{Pouquet1976} used the eddy-damped quasi-normal Markovian (EDQNM) approximation and obtained $-\delta_{ij}\tau\bar{\bmu\cdot\del\times\bmu}/3$ for the kinetic part of an isotropic $\alpha_{ij}$ pseudotensor, with $\tau$ being the correlation time of the turbulence.  Other analytical methods include the 
first order smoothing approximation (FOSA) \citep{KrauseRaedler1980,Moffatt1982}, 
minimal-$\tau$ approximation (MTA) \citep{BlackmanField2002}, 
second order correlation approximation (SOCA) \citep{RaedlerStepanov2006}, 
method of random waves \citep{Walder1980}, 
method of individual blobs \citep{Stix1983,Ferriere1992}, 
and path integral approaches \citep{Sokoloff2018}.

These analytical approaches show that the $\bm\alpha$ effect is related to helically correlated turbulent fields.
\citep[Note, however, that turbulent flows with no mean helicity may also produce an $\bm\alpha$-effect; see][]{Yousef2008,SridharSingh2010,SinghSridhar2011,SridharSingh2014,SinghJingade2015,Rasskazov2018pre}.

In many astrophysical systems, stratified turbulence combined with global rotation is believed to result in a net mean 
kinetic helicity, as shown in both 
analytical work \citep{SteenbeckKrauseRaedler1966, Moffatt1978, Ruediger1978, RaedlerStepanov2006}
and numerical simulations \citep{CattaneoHughes2006, HughesCattaneo2008, Kapyla2009, BrandenburgRaedler2012}.
In particular, \cite{SteenbeckKrauseRaedler1966} derived their famous result for the 
$\bm\alpha$ pseudotensor in a stratified rotator,
\beq
\alpha_{ij}=-\frac{16}{15}l^2 {\bm \Omega}\cdot\del\ln(\rho u_\text{rms})\delta_{ij}
\label{eqn:Steenbeck1966}
\eeq
where $l$ is the correlation length, $u_\text{rms}=\bar{u^2}^{1/2}$, and $\rho$ is the mean density.
Later investigations that allow  for anisotropy revealed that the $\delta_{ij}$ in Eqn. (\ref{eqn:Steenbeck1966}) should be replaced by the tensor $\hat{\bm\Omega}\cdot\hat{\bm D}\delta_{ij}-\mu(\hat{\Omega}_i\hat D_j+\hat\Omega_j\hat D_i)$, where $\hat{\bm\Omega}=\bm\Omega/\Omega$, $\hat{\bm D}$ is the unit vector in the direction of $\del(\rho u_\text{rms})$, and $\mu=11/24$ in \cite{Raedler2003} and $1/4$ in \cite{Brandenburg2013}.
There are also debates as to whether the momentum term in Eqn. (\ref{eqn:Steenbeck1966}) should be replaced by the kinetic energy $\rho\bar{u^2}$, e.g., \cite{Brandenburg2013}.
For incompressible flows, \cite{RaedlerStepanov2006} also gives a slightly different result.

In this work we develop a new approach to calculate the kinetic $\bm \alpha$-effect 
in  rotating stratified turbulence.
We do not propose a new closure model, but instead focus on calculating the correlation functions, i.e., we start from the result of the MTA closure \citep{BlackmanField2002},
\beq
\alpha_{ij}=\tau\epsilon_{imn}\bar{u_m\partial_j u_n} = \overline{a_{ij}},
\label{eqn:mta}
\eeq
where the latter equation defines the pre-averaged pseudotensor $a_{ij}=\tau\epsilon_{imn}{u_m\partial_j u_n}$.
We will calculate $\bm a$ using the velocity components of a single plume moving in a certain direction, subjected to the influence of the Coriolis force.
The Coriolis force is induced by the expansion or shrinkage of the plume when there is a change in its physical surrounding, typically due to  stratified mean fields, e.g., temperature or entropy gradient.
We do not specify the detailed modeling of the expansion or shrinkage but rather assume a power-law relation between the dimension of the plume and some relevant mean field, therefore making our scheme generally applicable.
The $\bm\alpha$ tensor is obtained by averaging $\bm a$ over all the possible orientations of the motion.

Our approach conceptually differers from previous analytic approaches 
which start formally from the MHD equations in the frame in which the
averaged transport tensors are sought without focusing on the local physics of individual turbulent  plumes.
We take a ``bottom-up,"  first principles approach,  first focusing on the physics of  local turbulent plume motions, and
then averaging over different realizations of plumes in an ensemble to compute the transport tensor. This helps to provide  new insight because the form of the  correlation tensor is  more transparently connected the local turbulent plume motions.

Some related  approaches have been developed in the literature.
 \cite{Ferriere1992,Ferriere1992b,Ferriere1993a} derived a $z$-dependent turbulent EMF and its corresponding $\bm\alpha$ tensor
  for galactic mean field dynamos, by considering the evolution of a uniform magnetic field due to a single supernova or superbubble explosion and then averaged over a spatial distribution of explosions.
The calculation of  \cite{Ferriere1992,Ferriere1992b,Ferriere1993a} does not include non-linear transfer terms in the MHD equations. 
Here we use the MTA closure and therefore  the nonlinearities have been 
 taken into account, at least approximately. In addition, our approach is independent of the energy source and the alignment of  global rotation and density gradient, and thus can be applied to systems other than galaxies.

For turbulence with a convective origin which is more relevant to stellar dynamos, \cite{DurneySpruit1979} assumed that the turbulent velocity is a superposition of eight components, and gave the expressions for the turbulent Reynolds stress and the heat flux.
Based on the same method, \cite{Durney1981} and \cite{DurneyRobinson1982} calculated the $\bm \alpha$ tensor for  rotating unstratified turbulence.

Subsequently, \cite{Stix1983} included  density stratification in his calculation of $\bm\alpha$, but his helicity density is calculated using the mean velocity, rather than 
taking the mean of the turbulent helicity density.
There are three major differences between our present work and that of \cite{Stix1983}:
(i) We do not invoke the full nonlinear equation of motion of $\bm u$, but rather  consider only the Coriolis force term which generates vorticity. This greatly simplifies the mathematics;
(ii) Following the definition of $\bm\alpha$ in Eqn. (\ref{eqn:mta}) ,  we take the average  only after $a_{ij}$ is obtained;
(iii) We do not make the {\it a priori} assumption that the $\bm \alpha$ tensor is isotropic.
The anisotropy arises naturally in our approach, and  the result agrees very well with previous much lengthier analytic derivations.

In Sec. \ref{sec:diag} we introduce the notion of the (pseudo)tensorial
 nature of $\bm\alpha$, and calculate  its  transformation properties  under a coordinate transformation.  In Sec. \ref{sec:construct} we describe our  physical model of $a_{ij}$  in detail and present our calculation of $\bm \alpha$ using the results of Sec. \ref{sec:diag}. Discussions and conclusions are given in Sec. \ref{sec:conclusion}.

\section{\bf\large\textalpha\ as a pseudotensor}
\label{sec:diag}
Given the expression of the $\bm \alpha$-pseudotensor Eqn. (\ref{eqn:mta}), it can be readily verified that
\beq
\epsilon_{ijk}\alpha_{jk}=\partial_k\bar{u_k u_i}-2\bar{u_i\partial_k u_k}.
\label{5}
\eeq
Via application of Reynolds rules of averaging for  homogeneous and incompressible flows, the RHS  of Eq. (\ref{5}) vanishes and the $\bm\alpha$-pseudotensor has to be symmetric, i.e., $\alpha_{ij}=\alpha_{ji}$.
Furthermore, even for inhomogeneous or compressible cases, $\bm \alpha$  can always be decomposed into its symmetric and anti-symmetric parts,
\beq
\bm\alpha=\bm\alpha^\text{S}+\bm\alpha^\text{A}
\eeq
where $\bm\alpha^\text{S,A}=(\bm\alpha\pm\bm\alpha^\text{T})/2$, and the superscript $\text T$ indicates transpose.
However, in the mean induction equation, the anti-symmetric part of $\bm\alpha$ multiplied by $\bm B$ has an advection-like behavior, and $\bm\alpha^\text{A}$ can be rewritten as a effective mean velocity field \citep{Charbonneau2014}.
We therefore focus on a symmetric $\bm \alpha$ pseudotensor in the rest of the paper.

The components of $\bm\alpha$ in a certain frame constitutes a real and symmetric matrix, and thus can be diagonalized by an orthogonal matrix constructed from its eigenvectors, e.g.,
\beq
\alpha'_{ij}=\left(S^{-1}\right)_{ia}\alpha_{ab}S_{bj}
\eeq
where $\alpha'_{ij}$ are the components of a diagonal matrix and $\bm S$ is real and orthogonal, $\bm S^{-1}=\bm S^\text{T}$.
In our formalism,   primed quantities are in the frame of diagonalization, and unprimed quantities are in the lab frame.
The determinant of $\bm S$ can be either $1$ or $-1$ since $1=\det(\bm S^\text{T}\bm S)=(\det\bm S)^2$.
The $\det\bm S=-1$ branch involves reflection operation in  configuration space and in order to remove this complexity, we can conveniently normalize $\bm S$ by its determinant and define $\bm R=\bm S/\det \bm S$.
Notably $\bm R$ can then be regarded as a rotation matrix, or an element of the group $SO(3)$.

Diagonalizing $\bm\alpha$ using $\bm R$  corresponds to a  
coordinate transformation of configuration space coordinates
by rotation of the axes about the origin. This can be readily seen by noting that  rotation by a matrix $\bm W\in SO(3)$  transforms  the coordinates and gradients as
\beq
x_i\to x'_i=W_{ij}x_j,\ 
\partial_i\to\partial'_i=W_{ij}\partial_j,
\eeq
and the components of vector and pseudo-vector quantities transform as
\beq
u_i\to u'_i=W_{ij}u_j,\ 
B_i\to B'_i=W_{ij}B_j.
\eeq
Since $\bm\alpha$ is a rank-2 (pseudo-)tensor, we have
\beq
\alpha_{ij}\to \alpha'_{ij}=W_{ia}W_{jb}\alpha_{ab}.
\eeq
Now if we identify $\bm W=\bm R^{-1}$, effectively we have rotated the system into a frame where the $\bm\alpha$ tensor is diagonalized.

We can now show that form of the induction equation is the same in the two frames. We write 
\begin{align}
\partial_t B'_i
=&\partial_t (W_{ij}B_j)\notag\\
=&W_{ij}\epsilon_{jkl}\partial_k(\alpha_{lm}B_m)\notag\\
=&W_{ij}W^{-1}_{jp}W^{-1}_{kq}W^{-1}_{lr}W^{-1}_{ks}W^{-1}_{lw}W^{-1}_{mu}W^{-1}_{mv}
\epsilon'_{pqr}\partial'_s(\alpha'_{wu}B'_v)\notag\\
=&\delta_{ip}\delta_{qs}\delta_{rw}\delta_{uv}
\epsilon'_{pqr}\partial'_s(\alpha'_{wu}B'_v)\notag\\
=&\epsilon'_{isw}\partial'_s(\alpha'_{wv}B'_v),
\end{align}
where we have used $(W^{-1})_{ij}=W_{ji}$.
The invariance of the form of the induction equation  between the lab and diagonalized frames  is essential for our method, given that boundary conditions are also accordingly transformed.
 It allows  us to study the  local physics of the system (e.g., energy and helicity growth) in the  diagonalized frame to construct $a_{ij}$, and then  transform to the lab frame and average  to get $\bm \alpha$.


\section{{\bf\large\textalpha} in a stratified rotator}
\label{sec:construct}
\subsection{Method overview}
Given the aforementioned  invariance of the induction equation, we now derive  the $\bm \alpha$  for turbulent stratified flows.
The strategy consists of three steps:
(i) Build an appropriate physical model for the flow in a {\it single} turbulent cell, and then calculate  $a_{ij}'=\tau\epsilon_{imn}'u_m'\partial_j' u_n'$ for this local turbulent cell in the  preferred, convenient frame $S'$.
 For example, 
$S'$ might be a frame in which the velocity components of the flow have simple forms;
(ii) Rotate $a_{ij}'$ back to the lab frame $S$ to get $a_{ij}$;
(iii) Obtain $\alpha_{ij}$ in the lab frame by averaging $a_{ij}$ over a distribution of rotational transformation matrices.
Equivalently, this means averaging over identical turbulent plumes with different orientations with respect to  fixed rotation and density stratification axes.

The theory behind step (iii) above warrants further explanation. 
First, by averaging over a distribution of $S'$ we are  ensemble averaging, since  $\bm a$ is fixed at a given time and location in a real system.
However, when the correlation length and time are small compared to the corresponding scales of the mean fields, the ensemble average will deviate only by a small   (but quantifiable  amount  \cite{ZhouBlackmanChamandy2018})
amount from  spatial and/or temporal averages.
Second, the realistic distribution of $S'$ is generally non-trivial, and is related to the actively studied question of building an accurate ensemble 
of turbulence (\citep{Kraichnan1973, Frisch1975, Shebalin2013}).
For simplicity, we  choose a uniform distribution of $S'$, and leave other choices for future work.

To understand the averaging procedure mathematically, consider a statistically homogeneous flow where $\bm a$
 in each turbulent cell is has a uniform value, and in a given local frame
  has components $a'_{ij}$.
To obtain the components of $\bm\alpha$ in the lab frame, $a'_{ij}$ of each cell has to be rotated accordingly, and the result averaged i.e.,
\beq
\alpha_{ij}=\sum_{p=1}^N f(p)R^{(p)}_{ia}R^{(p)}_{jb} a'_{ab}
\eeq
where $p$ labels different turbulent cells, $N$ is the total number of turbulent cells, $\bm R^{(p)}$ is the rotation matrix which transforms the $S'$ frame of the $i$th cell to the lab frame, and $f(p)$ is the fractional multiplicity of $\bm R^{(p)}$ in the ensemble.
In the continuum limit, the summation should be replaced by the integration of the parameters of the rotation group.

\subsection{Specific calculation}
We now employ the approach to derive the anisotropic $\alpha$ tensor in a rotating turbulent flow, and compare the result with previous analytical work.

\subsubsection{Kinetic helicity of plume in diagonalized frame}
\begin{figure}
	\centering
	\includegraphics[width=0.8\columnwidth]{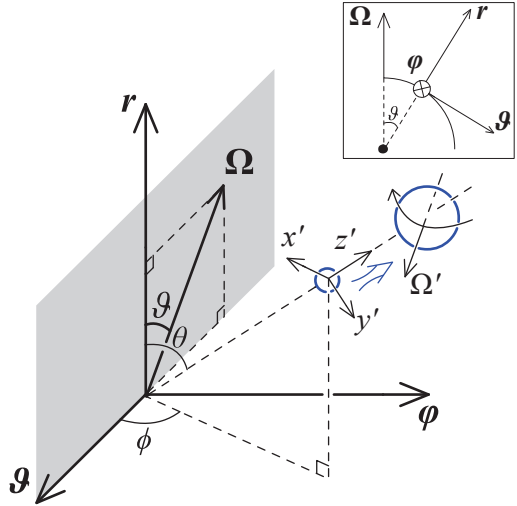}
	\caption{The coordinate systems used for calculations.
		The $\{\hat{\bm r},\hat{\bm\vartheta},\hat{\bm\varphi}\}$ is the spherical coordinate system of the background global rotator (e.g. star or galaxy) with global rotation vector $\hat{\bm\Omega}$, and ${\hat{\bm r}}$ points opposite to the density gradient.
		For galaxies $\hat{\bm\Omega}\cdot\hat{\bm r}=1$.
		The Cartesian frame $S$ is at colatitude and longitude $(\vartheta,\varphi)$ on the rotator, and its $\{\hat{\bm x},\hat{\bm y},\hat{\bm z}\}$ axes are parallel to $\{\hat{\bm\vartheta},\hat{\bm\varphi},\hat{\bm r}\}$ respectively.
		The $\{\hat{\bm x}',\hat{\bm y}',\hat{\bm z}'\}$  represents the local  Cartesian  coordinate system $S'$ of a local plume,  where $\hat{\bm z}'$ is along the direction $(\sin\theta\cos\phi,\sin\theta\sin\phi,\cos\theta)$ in $S$.
		Inset shows the 2-D projection of the system onto the plane spanned by ${\hat{\bm r}}$ and $\hat{\bm\Omega}$, for e.g. a star.
	}
	\label{fig:alphaFrame}
\end{figure}

We consider a plume  in the co-rotating frame to move 
 a distance $l$ over a turbulent correlation time at a constant velocity of magnitude $u=l/\tau$ before dissipating.
Motion perpendicular to the contours of mean density or turbulence intensity introduces a lateral expansion or shrinkage of the plume.
The Coriolis acceleration resulting from  this  lateral motion will  generate a net vorticity of the plume, parallel or anti-parallel to the direction of background global rotation.
When the magnetic field is flux-frozen into the conducting medium, the combined rise-and-twist motion of the flow produces a magnetic field component perpendicular to its original orientation.
For turbulence represented by an ensemble  of randomly oriented plumes,   
 the ensemble average effect gives rise to a mean kinetic $\bm\alpha$-effect.

To clarify the needed coordinate systems, consider a star with a global rotation $\bm\Omega$, as in the inset of Fig. \ref{fig:alphaFrame}.
The star is given a spherical coordinate system $(\hat{\bm r},\hat{\bm\vartheta},\hat{\bm\varphi})$.  The gradient of mean density or turbulence intensity is typically parallel or anti-parallel to $\hat{\bm r}$, along which the plume expands maximally.
In the local Cartesian frame $S$ at colatitude $\vartheta$ 
whose $\hat{\bm x}$, $\hat{\bm y}$ and $\hat{\bm z}$ axes are respectively $\hat{\bm\vartheta}$, $\hat{\bm\varphi}$ 
and $\hat{\bm r}$, $\bm\Omega$ has components $\Omega(-\sin\vartheta,0,\cos\vartheta)$.
For galaxies, $\vartheta=0$ and $\hat{\bm\Omega}\cdot\hat{\bm r}=1$ everywhere, but for stars, these relations may hold only at the poles.

Further, in the  Cartesian  frame $S$,
we parameterize the direction of the motion of a plume moving in a straight line as $\hat{\bm z}'=(\sin\theta\cos\phi,\sin\theta\sin\phi,\cos\theta)$.
Note that $(\vartheta,\varphi)$ and $(\theta,\phi)$ are two different sets of variables:
The former denotes the global colatitude and longitude (thus the ``location'' of $S$), whereas the latter denote the orientation of the plume in $S$ and are to be averaged over. 
We use  $S'$ to indicate the Cartesian frame whose vertical axis is $\hat{\bm z}'$.

We first study an ascending plume, and thus $\cos\theta>0$.
We assume that the plume  expands due to the background mean field gradient  (in $\hat{\bm r}$) and internal thermodynamics according to $L(r)\propto Q^q$, where $L$ is a characteristic length scale of the plume, 
$Q$ is some mean field (e.g., density, turbulent kinetic energy, etc.), and $q$ is an index determined by a turbulence model.
Assuming $Q$ has a large scale height compared to the correlation length, the relative change in $L$ due to the center-of-mass motion of the plume is
\beq
\frac{\Delta L}{L}
=\Delta\ln L
=q \Delta \ln Q
\approx q l\cos\theta \hat{\bm r}\cdot\del \ln Q,
\ \theta\in[0,\frac{\pi}{2}),
\label{eqn:DeltaR}
\eeq
where $l\cos\theta$ is the distance the blob travels along $\hat{\bm r}$.

The magnitude of the Coriolis force exerted on the plume during the expansion is $\sim 2\Omega\Delta L$, and thus it produces an angular velocity $-2\bm\Omega\Delta L/L$.
The component of the rotation vector perpendicular to $\hat{\bm z}'$ contributes $\mathcal{O}(\Omega^2)$ terms to $a'_{ij}$ and therefore will be neglected in the following calculations.
To find out the component along $\hat{\bm z}'$, note that
the cosine of the angle between $-\hat{\bm\Omega}$ and $\hat{\bm z'}$ is
\beq
\begin{array}{r}
	-\hat{\bm\Omega}\cdot\hat{\bm z}'
	=(\sin\vartheta,0,-\cos\vartheta)
	\cdot
	(\sin\theta\cos\phi,\sin\theta\sin\phi,\cos\theta)\\
	=-\left(\cos\theta\cos\vartheta-\sin\theta\sin\vartheta\cos\phi\right).
	\label{14}
\end{array}
\eeq
Thus in the $S'$ frame, to  $\mathcal{O}(\Omega)$ order the plume moves with velocity $u\hat{\bm z}'$
and  spins with angular velocity 
$-\Omega'\hat{\bm z}'$
where
\beq
\Omega'=2\Omega(\cos\theta\cos\vartheta-\sin\theta\sin\vartheta\cos\phi)\Delta L/L.
\label{eqn:omegaprime}
\eeq

Now the components  of $a'_{ij}$ can be  calculated for the plume using
$a'_{ij}=\epsilon_{imn}'{u_m'\partial_{j'} u_n'}$.
Writing out component by component and assuming $\partial_y' u_x'=-\partial_x' u_y'=-\Omega'$ and $\partial_z'=0$, it  follows that in the $S'$ frame the components of $\bm a$ reads
\beq
a'_{xx}=a'_{yy}=-\frac{1}{2}\tau h_\text{p},
\label{eqn:aturb}
\eeq
with other components all being zero, and
\beq
\begin{array}{r}
	h_\text{p}
	=-\frac{4q\Omega l^2\hat{\bm r}\cdot\del \ln Q}{\tau}
	\cos\theta
	\left(\cos\theta\cos\vartheta-\sin\theta\sin\vartheta\cos\phi\right).
\end{array}
\label{eqn:hb}
\eeq
The overall $\cos\theta$ factor assures that  both ascending and  descending plumes (the latter of which shrink and rotate oppositely to  ascending  plumes) are accounted for.
Eqns. (\ref{eqn:aturb}) and (\ref{eqn:hb}) are applicable in the slow rotation limit, characterized by low Coriolis number $\text{Co}=2\Omega\tau\lesssim1$.
For fast rotators, Taylor-Proudman-like columns form \citep[e.g., ][]{Brun2009} and higher order terms  in $\Omega$ must  be included.  

The trace of $\bm a$ is $-\tau h_\text{p}$, and we now show that $h_\text{p}$ is exactly the kinetic helicity of the plume.
The plume vorticity produced by the Coriolis force is
\beq
\del\times{\bm u}\simeq -2\Omega'\hat{\bm z}'.
\label{eqn:vorticity}
\eeq
The  extra factor of two results because $\del\times(\bm\Omega'\times\bm r)=2\bm\Omega'$ where $\bm\Omega'=\Omega'\hat{\bm z}'$.
The kinetic helicity of the plume is just
\beq
u\hat{\bm z}'\cdot\del\times\bm u=-2u\Omega'.
\label{eqn:udelu}
\eeq
Combining Eqns. (\ref{eqn:DeltaR}), (\ref{eqn:omegaprime}), (\ref{eqn:vorticity}) and (\ref{eqn:udelu}) yields Eqn. (\ref{eqn:hb}), confirming that the latter is indeed the kinetic helicity.


\subsubsection{Transformation to  lab frame and computation of \bf\large\textalpha}
To compute the components of $\bm a$ in the $S$ frame,
we make an orthogonal transformation using  the matrix $\bm R(\theta,\phi)$ which rotates the vector $(0,0,1)$ [representation for $\hat{\bm z}'$ in the $S'$ frame] to $(\sin\theta\cos\phi,\sin\theta\sin\phi,\cos\theta)$ [representation for $\hat{\bm z}'$ in the $S$ frame], namely:
\begin{align}
&\bm R(\theta,\phi)\notag\\
=&\begin{pmatrix}
	1-2\sin^2\frac{\theta}{2}\cos^2\phi & -\sin^2\frac{\theta}{2}\sin2\phi & \sin\theta\cos\phi \\
	-\sin^2\frac{\theta}{2}\sin2\phi & 1-2\sin^2\frac{\theta}{2}\sin^2\phi & \sin\theta\sin\phi \\
	-\sin\theta\cos\phi & -\sin\theta\sin\phi & \cos\theta
\end{pmatrix}.
\label{eqn:R}
\end{align}
Then $\bm \alpha$ in the $S$ frame is obtained by averaging $\bm a$ over all  possible orientations, namely
\beq
\alpha_{ij}
=\frac{1}{4\pi}
\int_0^{\pi}\sin\theta d\theta\int_0^{2\pi}d\phi\ 
R_{im}(\theta,\phi)
R_{jn}(\theta,\phi)
a'_{mn}.
\label{eqn:averaturb}
\eeq
In this  ensemble average, we  have assumed that the direction of  motion is distributed uniformly on a sphere, i.e.,
\beq
f(\theta,\phi)=\frac{1}{4\pi}\ \text{for}\ \theta\in[0,\pi)\ \text{and}\ \phi\in[0,2\pi),
\eeq
before it is perturbed by the Coriolis force, as expected for  slow rotators.
For fast rotators the plume would be more cylindrical \citep{Brun2009}.

Using Eqns. (\ref{eqn:aturb}), (\ref{eqn:hb}), (\ref{eqn:R}) and (\ref{eqn:averaturb}), we obtain
\beq
\bm\alpha
=\frac{8q}{15}
\Omega l^2 \hat{\bm r}\cdot\del\ln Q
\begin{pmatrix}
\cos\vartheta & 0 & \frac{1}{4}\sin\vartheta\\
0 & \cos\vartheta & 0\\
\frac{1}{4}\sin\vartheta & 0 & \frac{1}{2}\cos\vartheta
\end{pmatrix}
\eeq
in the $S$ frame.
More conveniently,  we can also write this as
\beq
\alpha_{ij}
=\frac{8q}{15}
\Omega l^2 \hat{\bm r}\cdot\del\ln Q
\left[
\delta_{ij}(\hat{\bm\Omega}\cdot\hat{\bm r})
-\frac{1}{4}(\hat\Omega_i\hat r_j+\hat\Omega_j\hat r_i)
\right],
\label{eqn:aijresult}
\eeq
independent of the choice of frame,
and it has trace
\beq
\tr\bm\alpha=\alpha_{ii}
=\frac{4q}{3}l^2
\left(\hat{\bm r}\cdot\bm{\Omega}\right)
\left(\hat{\bm r}\cdot\del\ln Q\right).
\eeq
This overall  factor of  $\tr\bm\alpha$ is determined by  the helicity gain due to the motion of the plume and in general would be determined by detailed modeling of turbulence and thermodynamics
of the plume motion. In this respect our approach accommodates different detailed models and thermodynamic assumptions for   plume motion.
For example, we may consider a simple scaling relation for the plume size   $L^3\propto \rho^{-1}\propto (\rho \bar{u^2})^{-3}$ where in the last step we have used $\rho u_\text{rms}^3=\text{constant}$ in \ mixing length theory.
This gives $Q=\rho \bar{u^2}$ and $q=-1$, and differs only slightly from previous  more rigorous calculations in the literature.
For example, \cite{SteenbeckKrauseRaedler1966} find $q=-2$ and $Q=\rho u_\text{rms}$; 
\cite{Raedler2003} find $q=2$ and $Q=\bar{u^2}$;
and \cite{Brandenburg2013} find $q=-2$ and $Q=\rho\bar{u^2}$.

However, our expression of the tensorial structure of $\bm\alpha$ [the part in the square bracket in Eqn. (\ref{eqn:aijresult})] is purely the result of averaging the motion of the plume over different directions, and only requires the assumption of  fully developed isotropic background turbulence.
That makes the  method  refreshingly  concise, and reproduces the correct results  even without the need to work in Fourier space.

\section{Conclusion}
\label{sec:conclusion}
Even when a closure is assumed, calculating turbulent transport coefficients for practical use in 
astrophysics can be  cumbersome  for stratified rotators because
of the anisotropies these quantities induce. 
Here we developed a mathematically simple and physically   transparent method 
and applied it to deriving the 
kinetic $\bm\alpha$ tensor  of  mean field dynamo theories.  In particular, we calculated the  motion of a single plume  in a  locally diagonalized frame, then ensemble averaged to get the lab frame 
mean transport coefficient.


We showed 
that our result Eqn. (\ref{eqn:aijresult}) matches very well with those from lengthier approaches \citep{Raedler2003, Brandenburg2013} that start with the same assumptions, namely:
(i) the only two non-trivial vectors in the problem are the rotation vector $\bm\Omega$ and the gradient of a mean field  flow property, 
Results are obtained to  first order in these two vectors.
(ii) The background turbulence without the rotation is statistically homogeneous, isotropic, and of high Reynolds number.
Previous approaches did  not exploit  simplifications that arise from
first working in the frame of a local plume and then transforming back to the lab frame.

Our approach is quasi-analytic;  we do  not solve the full Navier-Stokes equation.
More accurate results can be obtained by solving for the motion of the plume more rigorously, possibly with numerical simulations.
We have also assumed that the  direction of motion of the plume center of mass  has a uniform distribution over all spatial orientations, which may not be the case in a more realistic setting.

Our  approach  is not  restricted to the helicity tensor of  mean field dynamo theories. The method could be extended and applied to the calculation of other correlation functions  and transport coefficients.

\section*{Acknowledgments}
We acknowledge support from  grant NSF-AST-15156489,  and HZ  acknowledges support from Horton Fellowship from the Laboratory for Laser Energetics at U. Rochester. 


\bibliographystyle{mnras}
\bibliography{diagAlphabib}

\end{document}